# Nanoscale Graphene Electro-Optic Modulators

Zhaolin Lu* and Wangshi Zhao

*Microsystems Engineering, Kate Gleason College of Engineering,*

*Rochester Institute of Technology, Rochester, New York, 14623, USA*

*[*zhaolin.lu@rit.edu](*zhaolin.lu@rit.edu)

## Abstract

Research on graphene has revealed its unique optical properties[1], including strong coupling with light[2], high-speed operation[3], and gate-variable optical conductivity[4], which promise to satisfy the needs of future electro-optic (EO) modulators. In particular, recent work[5] demonstrated a broadband EO modulator based on the interband absorption of graphene with overall length only 40$\mu$m. However, compared with the size of on-chip electronic components it is still bulky. On-chip optical interconnects require EO modulation at the nanoscale. The key to achieve nanoscale graphene EO modulation is to greatly enhance light-graphene interaction based on novel waveguides and platforms. Herein, we present our recent exploration of graphene EO modulators based on graphene sandwiched in dielectric or plasmonic waveguides[6]. With a suitable gate voltage, the dielectric constant of graphene can be tuned to be very small due to the effect of intraband electronic transition, resulting in "graphene-slot waveguides" and greatly enhanced absorption modes. Up to 3-dB modulation depth can be achieved within 800nm long silicon waveguides, or 120nm long plasmonic waveguides based on small signal analysis. They have the advantages of nanoscale footprints, small insertion loss, low power consumption, and potentially ultrahigh speed, as well as being CMOS-compatible.

One of the most important devices in optoelectronic integrated circuits is the electro-optic (EO) modulator[7,8], which converts electronic signals into high bit-rate photonic data. Recent years have witnessed breakthroughs in the development of EO modulators[9-13]. However, the lack of ultrahigh-speed compact EO modulators remains a critical technical bottleneck impeding the wide deployment of the on-chip optical interconnects. Due to the poor electro-optic properties of regular materials [14, 15], a conventional EO modulator has a very large footprint[9,11,13, 16]. Employment of a high-Q resonator may significantly reduce the footprint, but it simultaneously decreases the operation bandwidth and thermal stability[10], which demand additional components to improve[17,18]. Hybrid of novel semiconductors[19-23] using sophisticated techniques may partially resolve these issues, but the involved waveguides are still tens to hundreds of micrometers long. PlasMOStors[24] can be very compact, but have inherently large insertion loss and limited operation speed. Recent research on graphene has provided unprecedented opportunities to meet the challenges.

Graphene[25,26] has attracted a great deal of interest because its exceptional electronic transport properties show great potential applications in the field of nanoelectronics[27] with the highest intrinsic mobility[28] and the largest current density at room temperature[29]. Graphene also has remarkable flexibility, robustness and environmental stability, as well as extraordinary thermal conductivity[30]. Equally outstanding are its optical properties, which have been used to develop light-emitters, ultrafast lasers, and photodetectors, as well as solar cells and touch screens[29]. As far as EO modulators concern, graphene is a single atom thick "film" with optical properties that are slightly dispersive and can be tuned in a large range at an ultrahigh speed through electrical gating–nearly an ideal electro-optic material.

We studied the optical properties based on the small signal analysis. Two absorption processes coexist in the light-graphene interaction, namely interband absorption and intraband absorption, which can be evaluated by a complex conductivity $\sigma_g = \sigma_{intra}(\omega,\mu_c,\Gamma,T) + \sigma_{inter}(\omega,\mu_c,\Gamma,T)$, depending on the light angular frequency $\omega$, chemical potential $\mu_c$, charged particle scattering rate $\Gamma$, and temperature $T$. The chemical potential $\mu_c$ can be controlled by electrical gating. Thus, the conductivity of graphene can be dynamically tuned by gate voltage $V_D$ in real time. Basically, when $\mu_c < \hbar\omega/2$ ($\hbar$ is the reduced Planck constant) interband absorption dominates and graphene becomes absorptive; otherwise, quite transparent. Electrically switching on/off graphene interband absorption plays a key role in the modulator reported in Ref. 5.

In our work, we found the intraband absorption can be equally important in a graphene absorption modulator. Based on the Kubo formula[31,32], we calculated the graphene conductivity at $T$=300K (scattering rate, $\hbar\Gamma$=5meV)[33]. Figure 1(a, b) plots the real and imaginary parts of the conductivity as a function of the chemical potential and wavelength in the near infrared regime. In particular, the real part of conductivity is very sensitive to chemical potential, for example at wavelength $\lambda_0$=1550nm, varying from nearly 60.85$\mu$S to 1.37$\mu$S when chemical potential rises from 0 to 0.6eV, as shown in Fig. 1(c). Figure 1(c) also shows how interband absorption and intraband absorption contribute to the graphene conductivity, respectively. Figure 1(d) plots the corresponding dielectric constant (real part, imaginary part, and magnitude),

$$\varepsilon_{eff}(\mu_c) = 1 - \frac{\sigma_v}{j\omega\varepsilon_0} = 1 - \frac{\sigma_g}{j\omega\varepsilon_0\Delta},$$ where $\Delta$=0.7nm is the effective thickness of graphene[5]. The dielectric constant of graphene varies from $\varepsilon_{eff}$(0eV)=0.985+j8.077 to $\varepsilon_{eff}$(0.6eV)=-2.508+j0.182 at $\lambda_0$=1.55$\mu$m. Note the sign of real part flips due to intraband absorption because the interband absorption and intraband absorption contribute the imaginary part of conductivity with different

signs as shown in Fig. 1(c). As a result, there is a dip in the curve of dielectric constant magnitude, where "metallic graphene" is transforming to "dielectric graphene" with Re$\{\varepsilon_{\text{eff}}\}$=0. In our case, the "transition chemical potential" is $\mu_t$=0.515 eV and $|\varepsilon_{\text{eff}}(\mu_t)|$= |-0.048+j0.323|=0.327, which means the magnitude varies $|\varepsilon_{\text{eff}}(0)|/|\varepsilon_{\text{eff}}(\mu_t)|$≈25 times! Note this "epsilon-near-zero" [34-36] effect can be seen almost in any material at its plasma frequency, for example Ag at $\lambda_0$=324 nm. The uniqueness of graphene lies in that its plasma frequency can be tuned by electrical gating.

The effect of dielectric constant change is not very manifest when graphene is placed on top of a dielectric waveguide. Based on the change of dielectric constant, we solved the transverse magnetic (TM) modes of graphene on a 250nm-by-600nm silicon waveguide with a 7nm $Al_2O_3$ buffer layer at $\lambda_0$=1.53$\mu$m when chemical potential is 0 and $\mu_t$, respectively. The effective indices are both 2.06, but the attenuation rates are significantly different, 0.134 dB/$\mu$m for $\mu_c$=0 and 0.044 dB/$\mu$m for $\mu_c$= $\mu_t$. The absorption can be further reduced when $\mu_c$ shifts from 0.515 eV to 0.6 eV. The resulting modulation, 0.09~0.13dB/$\mu$m, coincides the recent experimental work[5].

We find the absorption of a TM mode can be greatly enhanced when graphene is sandwiched inside the silicon waveguide, forming a "graphene-slot waveguide" as illustrated in Fig. 1(e). In a slot waveguide [37], the magnitude of transverse electric field $|E_y|$ is roughly *inversely* proportional to that of the dielectric constant. The power absorbed in a unit area,

$$p_d = \tfrac{1}{2}\text{Re}\{\sigma_g\}E^2 \propto \tfrac{1}{2}E \cdot \text{Im}\{\varepsilon_{\text{eff}}\}/|\varepsilon_{\text{eff}}|,$$

can be greatly enhanced at $\mu_c$= $\mu_t$ because (1) $|E_y|$ reaches its maximum and (2) $\text{Im}\{\varepsilon_{\text{eff}}\}/|\varepsilon_{\text{eff}}|$ nearly grows to its maximum at the same time as shown in Fig. 1(d). To verify this, we first consider the multilayer stack as illustrated in Fig. 1(e), where graphene is sandwiched in a silicon waveguide with a 10-nm $Si_3N_4$ buffer layer on each side. Based on the transfer matrix method, we find the optimal silicon thickness to enhance light absorption is about 150nm. Figure 1(f) plots the $|E_y|$ profiles at $\mu_c=0$ and $\mu_c=\mu_t$, respectively. The absorption is roughly proportional to $|E_y|$, with an enhancement about 25 times! In our case, $\mu_c=0$ is the transparence state, while $\mu_c=\mu_t$ is the absorption state, which are exactly *opposite* to the operation principle of the EO modulator reported in Ref. 5.

Once the configuration of the graphene-slot waveguide is optimized, we use a 3D mode solver to determine the optimal waveguide width based on the finite-difference time-domain (FDTD) method. Considering the fabrication tolerance, the optimal width of the waveguide is found to be 450nm. Figure 2(a) shows the mode profiles of the graphene-slot waveguide at different chemical potentials. There is only a slight shift in the effective index: 2.032 at $\mu_c=0$, and 2.034 at $\mu_c=\mu_t$. In contrast, there is a huge change in the waveguide attenuation. At $\mu_c=0$, the $|E_y|$ in the graphene is even lower than in the $Si_3N_4$ buffer layers, and the waveguide works at the low loss state with $\alpha_0=0.183$dB/$\mu$m; at $\mu_c=\mu_t$, the $|E_y|$ in the graphene is many times higher than in the $Si_3N_4$ buffer layers, and the waveguide works at the high absorption state with $\alpha_v=4.603$dB/$\mu$m. As a result, modulation depth 4.42dB/$\mu$m can be achieved, and 3dB-modulation depth only requires 679nm propagation distance! An 800-nm propagation distance results modulation depth 3.54dB. A graphene EO modulator can be made on the nanoscale! For the sake of easy fabrication, the silicon modulator can also take the form of an asymmetric slot waveguide as shown in Fig. 2(b). There is only a slight change in the performance.

Furthermore, recent work shows that highly confined modes can be achieved in plasmonic waveguides [38]. Based on nanoplasmonic platforms, the dimensions of a graphene modulator should be even smaller. Following the same approach, we investigated the interaction between graphene and various plasmonic modes. Figures 2(c,d) list the guided mode profiles, effective indices, and attenuation of graphene-slot waveguides based the metal-insulator-metal plasmonic platform. Due to the close interaction between metal and graphene, the chemical potential with highest absorption shifts to 0.518eV. Figures 2(e,f) list the mode calculation of graphene-slot waveguides based on the hybrid plasmonic platform. Although Au or Ag may decrease the metal absorption of the plasmonic waveguides, CMOS-compatible metal, Cu, is used in all plasmonic modulators, and its dielectric constant is assumed to be -67.86+j10.01. A 10-nm thick $Si_3N_4$ buffer layer is designed on each side of graphene for all plasmonic waveguides shown in Figs. 2(c-f). As can be seen in Fig. 2(d), a 3-dB (3.82dB at 1550nm) EO modulator can be made within 120nm using the metal strip plasmonic waveguide, where the attenuations are 6.76dB /$\mu$m at $\mu_c$=0, and 38.59dB /$\mu$m at $\mu_c$= 0.518eV.

To evaluate the insertion loss of the EO modulators, we performed 3D FDTD simulations with the smallest mesh size down to 0.35nm. In the simulations, we assume the modulators are embedded in the same waveguide as themselves except without the sandwiched graphene. We first simulated the modulator based on the silicon waveguide platform as shown in Fig. 3(a). The length of the graphene modulator is 800nm. Assume the thickness of the bottom silicon layer for electrical contact is negligible. Figures 3(b, c) show the power distribution in the waveguide at $\mu_c$=0 and $\mu_c$=0.515eV, respectively. Simulation results demonstrate that the overall throughput is 92.0% at $\mu_c$=0, and 42.5% at $\mu_c$= 0.515eV. Note that the insertion loss is only 0.36 dB (92.0%). The achievable modulation depth, 3.4 dB, is slightly smaller than the one predicted by the 3D mode solver.

We also simulated EO modulators based on the guided modes listed in Figs. 2(b-f). The results are similar as predicted in the mode solver. As one example, Figs. 3(e, f) show the simulation results at $\mu_c$=0 and $\mu_c$=0.518eV for the plasmonic modulator illustrated in Fig. 3(d). The overall throughput is 81.04% at $\mu_c=0$, and 36.92% at $\mu_c= 0.518$eV. Note that the overall length is only 120 nm while the modulation depth is 3.4 dB.

All numerical studies shown in Figs. 2 and 3 are performed at $\lambda_0=1550$nm for TM modes. On-chip optical interconnects require a broad bandwidth. Although the conductivity of graphene only weakly depends on the working frequency, the effective dielectric constant, $\varepsilon_{eff}(\mu_c)=1-\frac{\sigma_g}{j\omega\varepsilon_0\Delta}$, is a function of working frequency. As a result, in terms of dielectric constant, graphene is a dispersive medium. Nevertheless, we found the effect of dispersion is not so obvious. We studied the bandwidth of the EO modulators by solving the modes shown in Fig. 2 at different working wavelengths. Figure 4(a) shows the waveguide absorption as a function of wavelength in a silicon waveguide. As can be seen, the attenuation of the modulator at $\mu_c=0$ nearly remains a constant, 0.18-0.20 dB/$\mu$m, while the attenuation at $\mu_c=0.515$eV decreases when wavelength shift away from 1550nm. In particular, the attenuations are 4.44dB/$\mu$m, 4.60dB/$\mu$m, and 4.45dB/$\mu$m at 1545nm, 1550nm, and 1555nm, respectively. Wavelength spanning from 1545nm to 1555nm, or 1.25 THz bandwidth, only decreases modulation depth 0.16dB/$\mu$m. For our 800-nm silicon modulator, the decrease will be 0.14 dB. The prediction was further verified by 3D FDTD modeling. At $\mu_c=0$, the overall throughput is 92% for both 1545nm and 1555nm; at $\mu_c=0.515$eV, the overall throughput is 43.7% and 43.6%, for 1545nm and 1555nm, respectively. Thus, this modulator has a 3-dB bandwidth at least 1.25THz.

We also studied the bandwidth of our EO modulators based on plasmonic waveguides. Figure 4(b) shows the waveguide absorption as a function of wavelength in a metal strip plasmonic waveguide based on the 3D mode solver. As can be seen, when wavelength shifts ±5nm away from 1550nm, the attenuation decreases about 1.9dB/$\mu$m. Within 120nm, the modulation depth changes 0.23dB, from 3.82dB to 3.59dB. Thus, this EO modulator also allows for over a THz bandwidth. The calculation was also further verified by 3D FDTD modeling. At $\mu_c$=0, the overall throughput is 81% for both 1545nm and 1555nm; at $\mu_c$=0.518eV, the overall throughput is 37.84% and 37.58%, for 1545nm and 1555nm, respectively.

The modulator footprint mostly comes from the electrical contacts and the overall footprint can be made about 2~3μm$^2$ with the corresponding capacitance ~0.02 pF (the dielectric constant of Si$_3$N$_4$ is assumed to be 7.5). Note the magnitude of graphene dielectric constant is nearly stable between 0 and 0.4 eV as shown in Fig.1 (d). More accurately, significant output power decrease only occurs when the chemical potential varies from 0.445eV to 0.515eV as shown in Fig. 4(c,d). When projecting the chemical potential to gate voltage across a 10nm Si$_3$N$_4$ buffer layer, the gate voltage change $\Delta V$=(3.93-5.25)V. Thus, each bit only requires 0.12~0.13 pJ. Employment of doped graphene and a high-k (e.g. HfO$_2$) buffer layer can further decrease the power consumption.

Our modulators can potentially work at an ultra-high speed. Graphene has outstanding carrier mobility. In addition, intraband transition is much faster than interband transition[39]. The operation speed is mainly limited by the RC delay imposed by electric circuits. The sub-micrometer wide graphene may result in a very large resistance. Direct graphene-semiconductor contact may resolve this issue as shown in Fig. 5(a,b), and the RC delay can potentially decrease to several picoseconds. The resulting graphene-semiconductor Schottky barrier will be discussed in our future work.

We also considered the thermal transport in our modulators. Although graphene has a superior thermal conductivity, most heat still transfers through the buffer layers. In this case, we treat graphene as a thermal source. Assume the photonic signal power $P$=1mW (which is huge for telecommunications) and half is absorbed by graphene. Silicon nitride has a thermal conductivity $k$=29 W/mK. When applying the heat flux $\frac{P}{A} = \frac{0.5\text{mW}}{0.8\mu\text{m} \times 0.45\mu\text{m}}$ in the silicon waveguide-based modulator, the resulting temperature gradient in the Si$_3$N$_4$ buffer layer will be ~0.048°C/nm. The 10-nm buffer layer only results in temperature rise, 0.48°C.

All the theoretical analysis and numerical modeling in preceding text is based on the small optical signal assumption, i.e. the change of the graphene conductivity due to the absorption of light is negligible. Due to the extremely enhanced light absorption, saturable absorption and other nonlinear effects may become obvious when the signal power increases to some level. Actually, this nonlinear effect will become obvious when the pump signal is not so strong. According to our calculation, bias voltage $V_D$=5.3V will result in $\mu_c$=0.518eV across a 10-nm Si$_3$N$_4$ buffer layer with $N_s$=2.2×10$^{13}$ cm$^{-2}$. Absorption of light will give rise to excess carriers, which can be estimated by pump rate $R = \frac{P}{h\nu A}$ and carrier lifetime $\tau$ (~1ps for graphene), i.e., $\Delta N_s = \frac{P}{h\nu A}\tau$. For the modulator simulated in Fig. 3(d-f), $P$=0.1mW pump will result in $\Delta N_s$=3.3×10$^{12}$ cm$^{-2}$≈0.15$N_s$. Therefore, the modulator also provides us opportunities to study nonlinear optical effects at a low power level. One important application is all-optic modulators, where one weak optical signal ($\lambda_s$) may be switched on/off by another strong optical pump ($\lambda_p$) based on the graphene-slot plasmonic waveguide, where a DC bias voltage results in the maximum absorption of $\lambda_p$.

The proposed modulators can be fabricated with a series of standard semiconductor fabrication processes, such as thin film deposition, lithography, and etching. Graphene will be fabricated using the chemical vapor deposition (CVD) method[40,41]. Silicon nitride films will be deposited on top of the graphene films by plasma-enhanced chemical vapor deposition (PECVD) process, in which the plasma is maintained to be mild and low density, in order to avoid damage to the graphene film[42]. The bottom buffer layer will be either $Al_2O_3$ (or $HfO_2$) by atomic layer deposition or $Si_3N_4$ by PECVD.

To summarize, we studied the optical conductivity and dielectric constant of graphene under different chemical potentials in the near infrared regime. Due to the effect of intraband absorption, the magnitude of graphene dielectric constant (and hence the attenuation of a graphene-slot waveguide) can be dynamically tuned in a large range by electrical gating. We proposed and modeled a series of graphene EO modulators based on graphene-slot waveguides. Nanoscale graphene EO modulators can be developed based on both silicon and plasmonic platforms. These modulators promise to remove the technical bottleneck in on-chip optical interconnects with the advantages of nanoscale footprints, small insertion loss, low power consumption, and potential ultrahigh-speed, as well as being CMOS-compatible.

**Acknowledgements:** This material is based upon work supported in part by the U.S. Army under Award No. W911NF-10-1-0153 and the National Science Foundation under Award No. ECCS-1057381.

[22] H.-W. Chen, Y. H. Kuo, and J. E. Bowers, "25Gb/s hybrid silicon switch using a capacitively loaded traveling wave electrode," *Opt. Express* **18**, 1070 (2010).
[23] Y. Rong, *et al.* "Quantum-confined Stark effect in Ge/SiGe quantum wells on Si," *IEEE J. Sel. Top. Quant. Electron.* **16**, 85 (2010).
[24] J. A. Dionne, K. Diest, L. A. Sweatlock, and H. A. Atwater, "PlasMOStor: A Metal-Oxide-Si Field Effect Plasmonic Modulator," *Nano Lett.* **9**, 897-902 (2009).
[25] K. S. Novoselov, *et al.* "Electric field effect in atomically thin carbon films," *Science* **306**, 666–669 (2004).
[26] K. S. Novoselov, *et al.* "Two-dimensional gas of massless Dirac fermions in graphene," *Nature* **438**, 197 (2005).
[27] Y.-M. Lin, et al. "100-GHz Transistors from Wafer-Scale Epitaxial Graphene," *Science* **327**, 662 (2010).
[28] X. Du, I. Skachko, A. Barker, and E. Y. Andrei, "Approaching ballistic transport in suspended graphene," *Nat. Nanotechnol.* **3**, 491–495 (2008).
[29] A. K. Geim and K. S. Novoselov, "The rise of graphene," *Nat. Mater.* **6**, 183–191 (2007).
[30] A. A. Balandin, S. Ghosh, W. Bao, I. Calizo, D. Teweldebrhan, F. Miao, and C. N. Lau, "Superior thermal conductivity of single-layer graphene," *Nano Lett.* **8**, 902–907 (2008).
[31] V. P. Gusynin, S. G. Sharapov, and J. P. Carbotte, "Magneto-optical conductivity in graphene," *J. Phys.: Conens. Matter* **19**, 026222 (2007).
[32] G. W. Hanson, "Dyadic Green's functions and guided surface waves for a surface conductivity model of graphene," *J. Appl. Phys.* **103**, 064302 (2008).
[33] A. B. Kuzmenko, E. van Heumen, F. Carbone, and D. van der Marel, "Universal Optical Conductance of Graphite," *Phys. Rev. Lett.* **100**, 117401 (2008).
[34] M. Silveirinha and N. Engheta, "Tunneling of Electromagnetic Energy through Subwavelength Channels and Bends using ε-Near-Zero Materials," *Phys. Rev. Lett.* **97**, 157403 (2006).
[35] M. G. Silveirinha and N. Engheta, "Theory of supercoupling, squeezing wave energy, and field confinement in narrow channels and tight bends using ε-near-zero metamaterials," *Phys. Rev. B*, **76**, 245109 (2007).
[36] R. Liu, *et al.*, "Experimental Demonstration of Electromagnetic Tunneling Through an Epsilon-Near-Zero Metamaterial at Microwave Frequencies," *Phys. Rev. Lett.* **100,** 023903 (2008).
[37] Q. Xu, V. R. Almeida, and M. Lipson, "Experimental demonstration of guiding and confining light in nanometer-size low-refractive-index material," *Opt. Lett.* **29**, 1626 (2004).
[38] R. Yang, M. A. Abushagur, and Z. Lu, "Efficiently squeezing near infrared light into a 21nm-by-24nm nanospot," *Opt. Express* **16**, 20142 (2008).
[39] M. Breusing, C. Ropers, and T. Elsaesser, "Ultrafast Carrier Dynamics in Graphite," *Phys. Rev. Lett.* **102**, 086809 (2009).
[40] K. S. Kim, et al. "Large-scale pattern growth of graphene films for stretchable transparent electrodes," *Nature* **457**, 706–710 (2009).
[41] A. Reina, et al. "Large Area, Few-Layer Graphene Films on Arbitrary Substrates by Chemical Vapor Deposition", *Nano Lett.* **9**, 30–35 (2009).
[42] W. Zhu, D. Neumayer, V. Perebeinos, and P. Avouris, "Silicon Nitride Gate Dielectrics and Band Gap Engineering in Graphene Layers," *Nano Lett.* **10**, 3572–3576(2010).

**Figure captions**

Figure 1. (a) Real part and (b) imaginary part of the graphene conductivity as a function of chemical potential and wavelength ($T$=300K) based on the Kubo formula. (c) The graphene conductivity (real part and imaginary part), by interband transition and intraband transition, as the function of chemical potential at $\lambda_0$=1550nm. (d) The effective dielectric constant (real part, imaginary part, and magnitude) as the function of chemical potential at $\lambda_0$=1550nm. (e) The illustration of a 2D "graphene-slot waveguide" with a 10 nm thick $Si_3N_4$ buffer layer on each side of graphene. (f) The plots of the transverse electric field magnitude across the waveguide at $\mu_c$=0 and $\mu_c= \mu_t$, respectively.

Figure 2. The transverse electric field profiles, effective indices, and propagation loss for different graphene-slot waveguides at $\mu_c$=0 and $\mu_c= \mu_t$, respectively: (a) in a dielectric waveguide (Si waveguide is 450nm wide and 150nm thick for each layer); (b) in a dielectric strip waveguide (strip Si waveguide is 450nm wide and 150nm thick for each layer); (c) in a metal-insulator-metal waveguide (waveguide is 200nm wide); (d) in a metal strip waveguide (strip metal is 200nm wide); (e, f) in photonic-plasmonic hybrid waveguides (waveguide is 400nm wide in (e) and 200nm wide in (f), Si layer is 130nm thick for both structures). The refractive indices of Si, $Si_3N_4$, and $SiO_2$ are assumed to be 3.47, 1.98, and 1.44, respectively.

Figure 3. (a) The illustration of a graphene EO modulator based on a silicon waveguide. (b,c) The 3D simulation of light propagation between a silicon waveguide and the EO modulator at $\mu_c$=0 and $\mu_c= \mu_t$, respectively. (d) The illustration of a graphene EO modulator based on a metal strip plasmonic waveguide. (e,f) The 3D simulation of light propagation between a metal strip plasmonic waveguide and the EO modulator at $\mu_c$=0 and $\mu_c= \mu_t$, respectively.

Figure 4. The attenuation of graphene-slot waveguides as a function of working wavelength at $\mu_c$=0 and $\mu_c= \mu_t$, respectively: (a) in a silicon waveguide; (b) in a metal strip waveguide. The attenuation of graphene-slot waveguides as a function of chemical potential and gate voltage at $\lambda_0$=1550nm: (c) in a silicon waveguide (450nm wide and 150nm thick for each layer); (d) in a metal strip plasmonic waveguide (strip metal is 200nm wide).

Figure 5. The illustration of nanoscale graphene modulators containing direct graphene-semiconductor contacts based on (a) dielectric strip waveguide, and (b) metal strip waveguide.

**Figures**

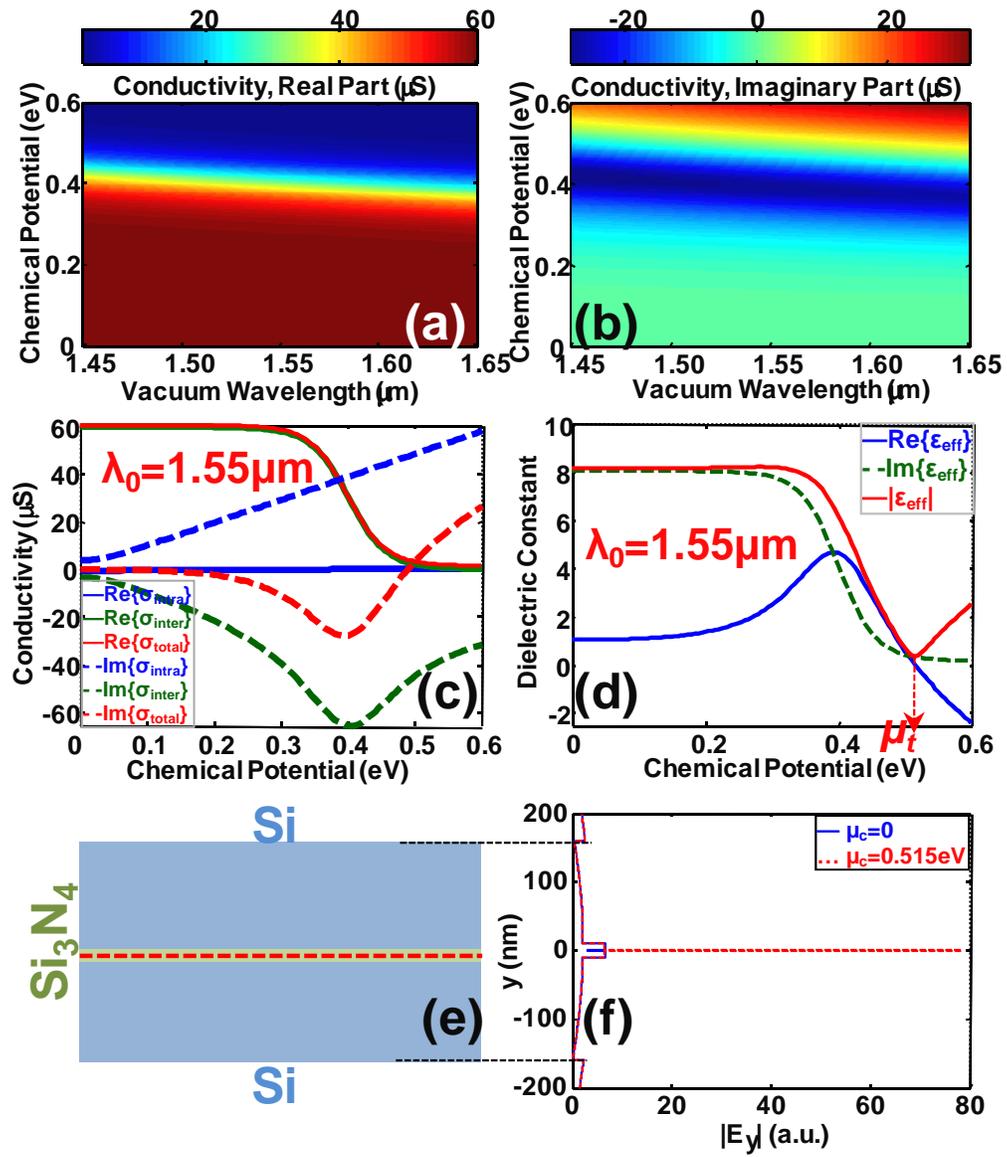

**Figure 1.**

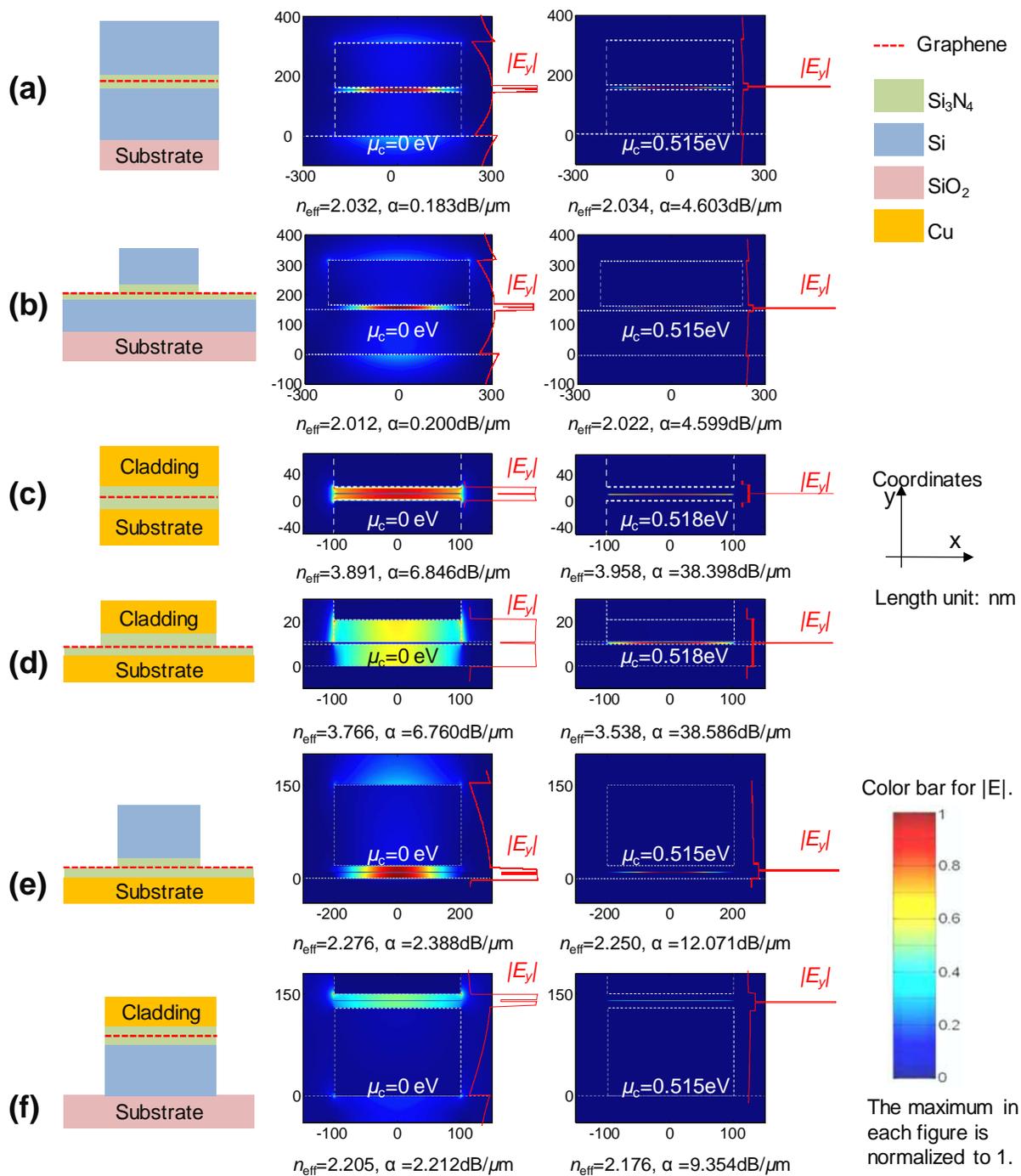

Figure 2.

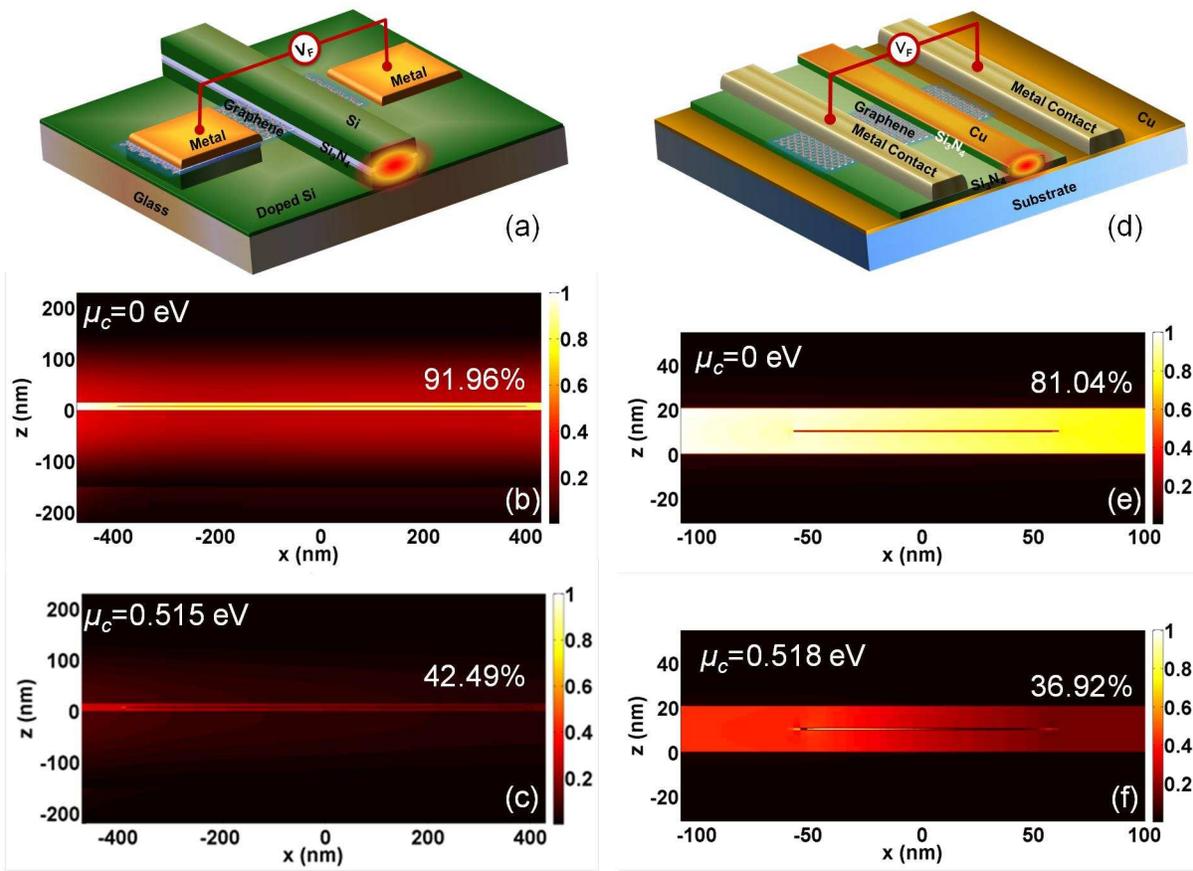

**Figure 3.**

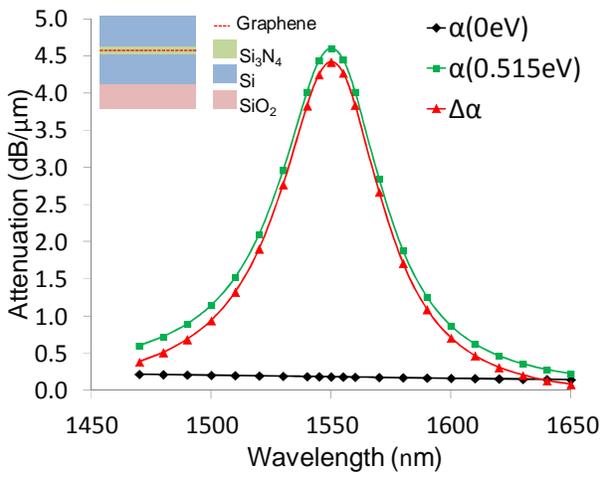
(a)

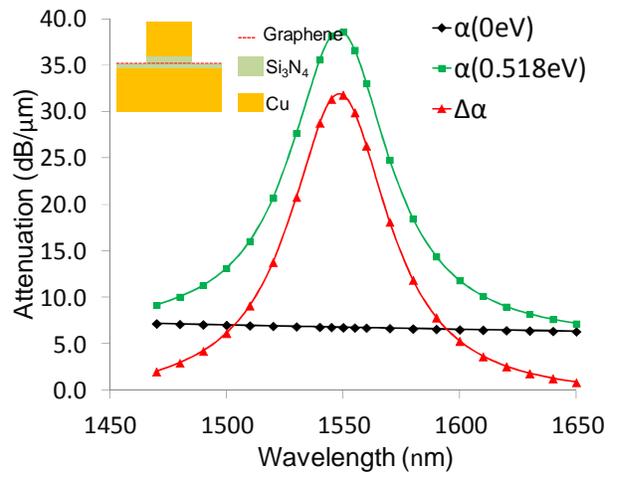
(b)

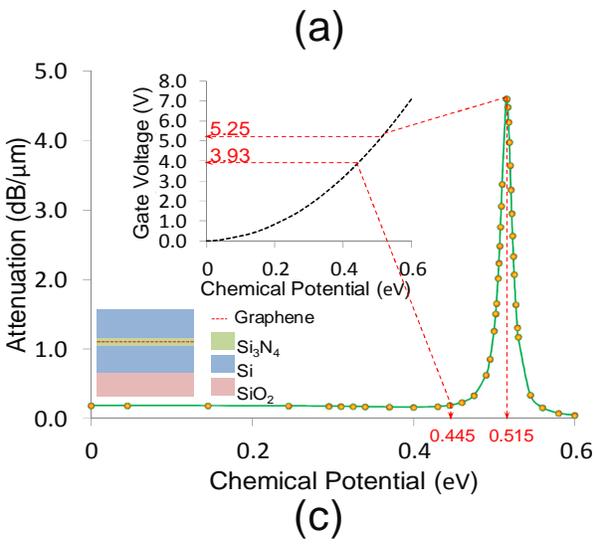
(c)

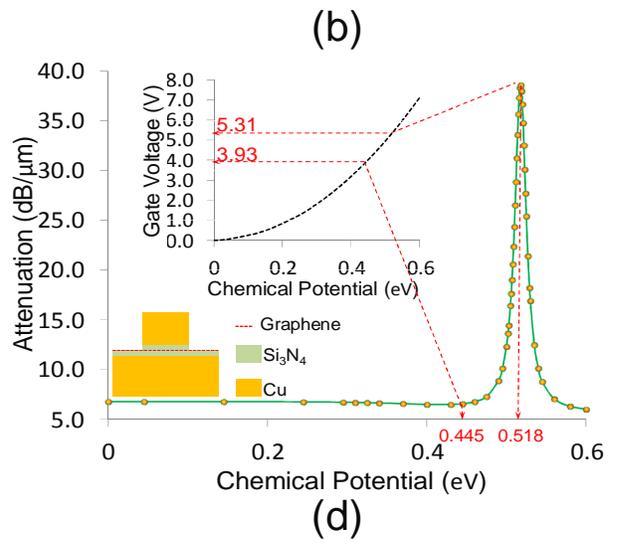
(d)

**Figure 4.**

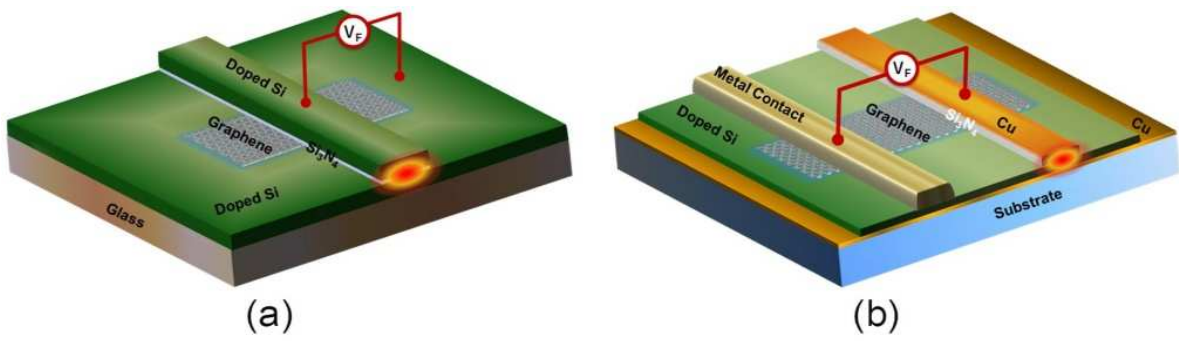

**Figure 5.**